\documentclass[sigconf]{acmart}

\usepackage{chngcntr}
\usepackage{graphicx}
\usepackage{amsmath}
\usepackage{subcaption}
\usepackage{amsfonts}
\usepackage{algorithmic}
\usepackage{soul}
\usepackage{url}
\usepackage{balance}
\usepackage{multirow, multicol, makecell}
\usepackage{url} 
\usepackage{xspace}
\usepackage{framed}
\usepackage{enumitem} 
\usepackage[breakable,skins]{tcolorbox}

\title{\textit{``Silent Is Not Actually Silent''}: An Investigation of Toxicity on Bug Report Discussion}

\author{Mia Mohammad Imran*}
\affiliation{%
  \institution{Missouri University of Science and Technology}
  \city{Rolla, Missouri}
  \country{USA}
}
\email{imranm@mst.edu}

\author{Jaydeb Sarker*}
\affiliation{%
  \institution{University of Nebraska  Omaha}
  \city{Omaha, Nebraska}
  \country{USA}
}
\email{jsarker@unomaha.edu}
\thanks{* These authors contributed equally to this work.}

\begin{abstract}
Toxicity in bug report discussions poses significant challenges to the collaborative dynamics of open-source software development. Bug reports are crucial for identifying and resolving defects, yet their inherently problem-focused nature and emotionally charged context make them susceptible to toxic interactions. This study explores toxicity in GitHub bug reports through a qualitative analysis of 203 bug threads, including 81 toxic ones. Our findings reveal that toxicity frequently arises from misaligned perceptions of bug severity and priority, unresolved frustrations with tools, and lapses in professional communication. These toxic interactions not only derail productive discussions but also reduce the likelihood of actionable outcomes, such as linking issues with pull requests.  Our preliminary findings offer actionable recommendations to improve bug resolution by mitigating toxicity. 

\end{abstract}



\copyrightyear{2025}
\acmYear{2025}
\setcopyright{cc}
\setcctype{by}
\acmConference[FSE Companion '25]{33rd ACM International Conference on the
Foundations of Software Engineering}{June 23--28, 2025}{Trondheim, Norway}
\acmBooktitle{33rd ACM International Conference on the Foundations of
Software Engineering (FSE Companion '25), June 23--28, 2025, Trondheim,
Norway}\acmDOI{10.1145/3696630.3728502}
\acmISBN{979-8-4007-1276-0/2025/06}

\begin{document}
\maketitle

\begin{sloppypar}

\section{Introduction}
\label{sec:intro}











Bug reports serve as a means of communication between users and developers~\cite{ko2011design, ko2010power}. They enable the identification and resolution of software defects, driving improvements in functionality and reliability~\cite{zimmermann2010makes}. Beyond their technical utility, bug reports help foster collaboration and build trust among community members~\cite{liang2022understanding, guizani2023rules}. They provide a structured medium for discussing software issues, proposing solutions, assessing the severity of problems, and prioritizing tasks~\cite{chaparro2017detecting, wang2022clebpi, song2022toward, gomes2019bug, zimmermann2010makes}.

Bug reports require precise and professional communication, as they often contain technical descriptions and demand clear, constructive feedback to facilitate issue resolution. However, bug reports are not without challenges. 
Zimmermann \textit{et} al. noted that \textit{``there is a mismatch between what developers consider most helpful and what users provide''} in bug reports~\cite{zimmermann2010makes}. Prior research has revealed that bug reports are frequently incomplete and ambiguous, with their quality varying~\cite {CHAUHAN2024107549practitioners, yusop2016reporting, davies2014s}, which often leads to bug reports not being addressed. Ko \textit{et} al. reported that users became dissatisfied when their bug reports were not addressed~\cite{ko2010power}.

\begin{figure}[tb]
    \centering
    \includegraphics[width=1.0\linewidth]{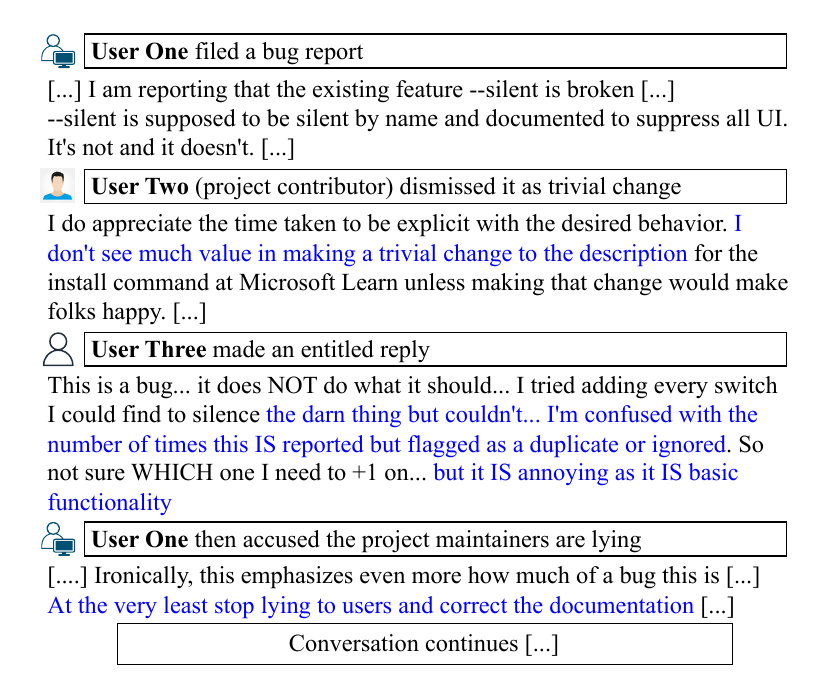}
    \caption{Toxic Interaction in Bug Report Discussion (\textcolor{blue}{blue} marked text shows toxic phrase)}
   \vspace{-4pt}
    \label{fig:motivating-example}
\end{figure}

These frustrations can manifest in negative ways, including toxic interactions and other antisocial behaviors. Users may express impatience, dissatisfaction, or hostility when their concerns are dismissed or unresolved. Figure~\ref{fig:motivating-example} illustrates a real-world bug report where the project maintainer dismissed a bug as a trivial change, sparking toxic comments.
Toxicity in bug reports disrupts communication, hampers collaboration, and may discourage users from participating in or continuing to use the project. 
For example, in this bug report~\cite{kitty-issue-7453}, the problem was not addressed, and the user was mocked by the owner. Subsequently, the user commented about switching to a different technology. In another thread~\cite{podman-issue-22363}, a user referenced market competition when a bug was not addressed (\textit{``I really don't mean to be total a*s about it but that's not how you win market share.''}).
These incidents show how toxicity in bug reports can escalate, disrupt productive discourse, and ultimately derail the collaborative efforts necessary for resolving software issues.

While prior research has investigated toxicity in various broader Software Engineering contexts, including GitHub pull requests (PRs) and issues threads~\cite{sarker2023automated, miller2022did, sarker2020benchmark, ferreira2022heated, sarker2023toxispanse, mishra2024exploring, imran2025understanding, sarker2025landscape}, bug reports remain an underexplored yet critical subset. Unlike other interactions, bug reports frequently highlight deficiencies in software, which can evoke emotions, such as frustration~\cite{zimmermann2010makes, imran2024uncovering}, and lead to toxicity~\cite{balavzikova2019real}.

To address the gap in analyzing toxicity on bug reports, we have conducted a qualitative study on how toxicity manifests in bug reports and impacts the developers' bug resolution. To achieve this goal, we manually investigated 203 bug threads, including 81 with toxic interactions, which were selected using a stratified sampling strategy. After a qualitative exploration, we have found several notable findings:

\begin{itemize}[leftmargin=1em]

    \item Toxicity arose when users perceived a disconnect between bug severity and maintainers' prioritization decisions.
    
    
    \item Toxic interactions hindered collaboration, with unresolved threads less likely to yield pull requests or fixes.
    \item Toxicity extended beyond interpersonal interactions and included negativity toward technology. 
\end{itemize}

\section{Methodology}
\label{sec:method}

In the following subsections, we have provided a brief description of our research methodology.

\vspace{-6pt}
\subsection{Dataset Creation and Labeling}
\subsubsection{Data Mining}

We selected GitHub-based repositories for our study as it is the most popular open-source software project hosting platform~\cite{octoverse-2023, fang2022slick}. 
Primarily, we selected the repositories from GitHub for our analysis based on three criteria: i)  at least 1000 stars to ensure sufficient community interest; ii) active in 2024 (up to September 5, 2024); and iii) active moderation was another key criterion. We defined it as evidence of maintenance and oversight, such as locking issues or the existence of a Code of Conduct included in the repository. After applying these criteria, we found 257 GitHub projects. 




Further, we sampled from the repositories chosen based on the following criteria:
i) all comments within threads must be in English to ensure consistency and accessibility; ii) created in 2024; iii) the issue/PR threads have at least 2 comments; and iv) each thread associated with a GitHub label related to bugs, such as \texttt{bug}, \texttt{defeat}, \texttt{crash}, \texttt{triage} or \texttt{fix}~\cite{kallis2021predicting, imran2021automatically}. Finally, we ended up with a total of 100 repositories and 8,723 threads with 91,929 comments. 


\subsubsection{Toxicity Classification}

\begin{table}[]
    \centering
    \footnotesize
    \caption{Data Sampling}
    \begin{tabular}{|c|c|c|c|}
\hline
Issue Thread & Count & ToxiCR ($\geq 0.5$) & LLaMA ($\geq 0.7$)  \\ \hline

 Total Thread & 8,723 & 861 &  789 \\ \hline
 Selected Thread & 203 & 186  & 165\\ \hline

 Labeled as Toxic & 81 & 80 & 59 \\ \hline
 
\end{tabular}

    \label{tab:data_sapmpling}
\end{table}

Since toxic conversations are rare in SE texts~\cite{miller2022did, sarker2020benchmark}, we employed automated tools to detect toxicity in the selected comments on our dataset. To improve the accuracy and reliability of our toxicity detection process, we utilized two toxicity detection methods. 
First, we selected ToxiCR~\cite{sarker2023automated}, a state-of-the-art SE-specific toxicity detector. Second, we applied the LLaMA model to toxicity detection for our dataset. We provided the details in the following:

\textbf{ToxiCR:} A cutting-edge toxicity detection tool referred to as ToxiCR~\cite{sarker2023automated} has been specifically designed to detect toxic comments in code reviews, which also got attention from SE researchers~\cite{rahman2024words}. Since this tool is publicly available with a $95.8\%$ accuracy and $88.9\%$ recall, we chose this toxicity detector for our analysis. 


\textbf{LLaMA Model:} Since its inception, LLaMA~\cite{touvron2023llama} has become a popular LLM used for various affective analysis tasks, including toxicity detection~\cite{koh2024can, inan2023llama}. We chose the LLaMA model as it is an open-source model, unlike proprietary models like GPT-4.  We utilized~\textit{LLaMA-3.1-70B} version. We asked the model to predict the likelihood of the conversation being toxic, given a thread on a scale of 0 to 1. We utilized the definitions of toxicity provided by Miller \textit{et} al.~\cite{miller2022did} and incivility characteristics defined by Ferriera~\textit{et} al.~\cite{ferreira2022heated} and Ehsani \textit{et} al.~\cite{ehsani2023exploring}. The detailed prompt is included in the replication package~\cite{dataset}.

\textbf{Dataset Sampling: } According to the authors of ToxiCR~\cite{sarker2023automated}, a text would be marked as toxic if it scores $\geq 0.5$. Hence, we set a threshold of ToxiCR as $\geq 0.5$ and LLaMA toxicity detection as $\geq 0.7$. We found 148 threads using these thresholds where both ToxiCR and LLaMA detected at least one toxic conversation. To expand our dataset size, we further sampled 55 threads where either ToxiCR or LLaMA has a toxicity score $\geq 0.99$. Thus, we ended up with 203 threads, forming the basis of our toxicity analysis in bug reports. Table~\ref{tab:data_sapmpling} shows the dataset selection for our analysis. 

\textbf{Manual Labeling:}  Toxicity is a complex phenomenon and varies according to different diverse demographic factors~\cite{kumar2021designing}. Moreover, the perception of toxicity differs in the SE domain from other social media platforms~\cite{miller2022did}. 
Therefore, two raters (authors of this paper) carefully reviewed prior literature on toxicity and other antisocial behaviors~\cite{sarker2020benchmark, miller2022did, ferreira2021shut} in SE to reduce bias during manual labeling. They set five rounds of discussion sessions and collaboratively labeled the 203 threads~\cite{saldana2021coding}. They reviewed each thread, carefully read the comments, and marked the thread as toxic if at least one comment was labeled as toxic by two raters. At the end of this process, we found 81 toxic threads.




\vspace{-6pt}
\subsection{Qualitative Analysis Of Bug Threads}

Inspired by recent works with toxicity and incivility analysis in the SE domain, we have conducted a qualitative analysis of selected toxic threads~\cite{miller2022did, ferreira2022heated}. We have manually analyzed which types of toxicity are more likely to occur in bug report discussions. We have found a total of 11 sub-categories of toxicity in our dataset. Table~\ref{tab:toxicity_types_impacts} shows our findings and subcategories of toxicity in GitHub bug discussions. We have also provided the mapping of subcategories in the second column of table~\ref{tab:toxicity_types_impacts} where two of them are not previously reported in any SE studies -- `\textit{Toxicity Towards Tools}' and `\textit{Profane Technology Naming}'. Developers sometimes get frustrated with using some existing tools and use toxic comments toward that technology, which is referred to as the `\textit{Toxicity Towards Tools}' category. Furthermore, we have found that the use of profane naming, including variable name method name, is an unprofessional practice, and we have defined that category as `\textit{Profane Technology Naming}'. Further, two of our raters manually went through all the toxic threads and looked into the following factors:

\begin{itemize}[leftmargin=1em]
    \item \textbf{Bug Resolved}: Whether the bug was resolved after the existence of toxic comments in the thread. 

    \item \textbf{Authors of Toxicity}: We examined the authorship characteristics of toxic comments based on their membership within the project. GitHub provides the labels of the authors~\cite{githubauthorship}. Authors were classified as internal participants if they were contributors, collaborators, members, and owners. Otherwise, they were categorized as external participants. 
    
    \item \textbf{Toxic Reply}: We considered a toxic reply when a developer responded with toxic text after receiving a toxic comment from peers within the same thread.

\end{itemize}

\begin{table*}[]
\vspace{-4pt}
    \centering
    
    \caption{Subcategories of Toxicity and Associated Factors (E = External participants, I = Internal participants)}
    \begin{tabular}{|p{2.5cm}|p{1cm}|p{7.6cm}|r|r|r|}
\hline

 Type & Mapping & Example & Count$\ddag$ &  Resolved &  Toxic Reply\\ \hline

 Vulgarity & \cite{ferreira2021shut, miller2022did, sarker2020benchmark} & \textcolor{red} {Why is it so hard to install this thing? It has been 3 f**ing hours already!!!} & 19 (E: 16, I: 3)  &  $31.6\%$ & 3 \\ \hline
 
 Insult & \cite{miller2022did, sarker2020benchmark}& \textcolor{red} {It's really very annoying to debug [...] Is anybody alive here? Do you need reproduction? Or what?} & 18 (E: 14, I: 4)  &  $38.9\%$ & 9 \\ \hline

 Unprofessional & \cite{miller2022did, sarker2020benchmark}& 
  \textcolor{red} {Darn it, now I can't reproduce after a restart of the window manager.}& 17 (E:10, I:7) &  $70.5\%$  & 0\\ \hline

 Entitlement & \cite{miller2022did} &\textcolor{red} {IMO this is a disappointing outcome. No reliable guideline for anything but dbus/now.} & 14 (E:9, I:5) &  $35.5\%$ & 9  \\ \hline

 Toxicity Towards Tools & New & \textcolor{red}{This means that Telegram Desktop [...] 5.0.1 and 5.0.2 contains shameful destructive bug} & 11 (E: 8, I: 3)  & $54.5\%$ & 1  \\ \hline

 Mocking & \cite{ferreira2021shut} &   \textcolor{red}{This isn't Windows compatible at all, are you joking?} & 4 (E: 4)  & $50\%$ & 2 \\ \hline

 Frustration & \cite{ferreira2021shut} & \textcolor{red}{why the hell is no error message being reported, that is some real bug within podman. I still need to debug this.} & 3 (E: 1, I: 2)  & $33.33\%$ & 1\\ \hline

 Profane Technology Naming & New & \textcolor{red}{plugins=(git.... thefuck.....)} & 2 (E:2)  & $0\%$ & 0\\ \hline
 
 Identity Attack & \cite{sarker2020benchmark} & \textcolor{red}{And then we don't have to pay Elon to message you...} & 1 (E:1)  & $100\%$ &\\ \hline

 Threat & \cite{ferreira2021shut, sarker2020benchmark} & \textcolor{red}{I think you're just trolling, so enjoy the ban and learn some manners.} & 1 (I:1) & $100\%$ & 1\\ \hline

Arrogant & \cite{miller2022did}& \textcolor{red}{Whatever this function is supposed to achieve, it's very weird and messed up. Wouldn't it make more sense to fix this function like I suggested..} & 2 (I:2)  & $100\%$ & 0\\ \hline

\multicolumn{6}{l}{$\ddag$ -\textit{Since a text can fall into multiple sub-categories, the total count across all categories exceeds the sample size}}

\end{tabular}
    \label{tab:toxicity_types_impacts}
 
\end{table*}
   \vspace{-4pt}

\section{Early Findings and Observations}
\label{sec:esults}


In this section, we present the preliminary findings from our qualitative analysis of 81 toxic issue threads. Several noteworthy observations emerged, including those summarized in Table~\ref{tab:toxicity_types_impacts} and found in our analysis, which are discussed in detail below. 

\smallskip
\noindent
\textbf{I. Subcategories of Toxicity}:  Table~\ref{tab:toxicity_types_impacts} shows 11 toxic categories that we found in our sample. Similar to prior studies on open source projects, we have found that vulgarity, insult, unprofessional, and entitlement are the most common categories encompassing $80\%$ of the samples.    
In $13.58\%$ cases, we observed toxicity directed toward tools, often characterized by negative or derogatory descriptions. While these comments were not aimed at individuals, they contribute to an unprofessional atmosphere and reveal deeper frustrations or dissatisfaction with the tools in question. For example, a user talked about missing docs in NextAuth by remarking - ``\textcolor{red}{Some examples would be awesome, docs or otherwise. The amount of hours wasted on nextauth is mindboggling. [...]}''~\cite{nextauthjs-next-auth-issues-10633}. We observed that 2 of our threads contain tool's naming containing profanity (e.g., \textcolor{red}{thef*ck}). While these may not be intended to be toxic, they are inappropriate in a professional environment.


\smallskip
\noindent
\textbf{II. Bug Report Authors and External Participants Use More Toxicity}: We observed that 41/81 (50.61\%) toxic comments were authored by the same individual who created the bug report thread. This suggests that bug report authors are more likely to initiate toxic comments than other discussion participants. This behavior could stem from their heightened emotional investment in the issue, frustration over perceived neglect, or dissatisfaction with the responses they receive. 
Additionally, 56/81 (69.31\%) first toxic comment authors were external participants. Though we worked with bug report threads, our results support the findings of a prior study~\cite{miller2022did}. An external participant may have many expectations and do not need to follow the Code of Conduct, which may trigger the use of more toxic comments than an internal participant. 
However, the internal participants authored 30.69\% of toxic comments, which is not non-trivial, which indicates that active contributors could also be significant contributors to toxic discourse.

\smallskip
\noindent
\textbf{III. Bug Severity and Bug Priority}: According to our observation, toxicity emerged when users perceived a bug as critical or a ``showstopper,'' while maintainers either dismissed it or failed to address it promptly. This disconnect between users’ perception of severity and maintainers’ prioritization sparked frustration and led to toxic interactions. We found 17 threads where this happened. 
For example, in one bug report~\cite{facebook-react-issue-28584}, a user commented - \textcolor{red}{``this is such an annoying bug, this has existed for nearly half a decade; but the people working on this are more obsessed with useless things like server components.''}. Users occasionally questioned why certain bugs were not prioritized. For instance, one user~\cite{strapi-strapi-issue-19339}, commented - \textcolor{red}{``yes its frickin annoying. my whole project is on hold now.''}


\smallskip
\noindent
\textbf{IV. Bug Resolution Rate:} Unsurprisingly, our analysis suggests that the presence of toxic comments negatively impacts the bug resolution rate. Among our sample, we found that a total of 36 bug threads ($45\%$) were resolved. However, in threads containing vulgarity, insults, entitlement, and expressions of frustration, the bug resolution rate drops to approximately one-third, highlighting the detrimental effect of toxicity on resolution outcomes. This finding would help the project managers mitigate specific toxicity subcategories in the project discussions.

\smallskip
\noindent
\textbf{V. Absence of PRs from Toxic Threads}: We observed that only 2 bug threads were PRs. This indicates that when a pull request addresses a bug, it is less likely to become toxic. 
One possible explanation is that pull requests are often tied to tangible progress and positive outcomes, such as bug fixes, which tend to elicit satisfaction and collaborative behavior rather than conflict~\cite{zhang2022pull, winter2022developers}.

\smallskip
\noindent
\textbf{VI. Issue Bug Threads Linkage with PRs}: GitHub bug reports that require \textit{fixes} are often associated with pull requests~\cite{de2024revealing, li2018issue}. This indicates the bug reports are addressed in those issues. We found that 23/79 (29.11\%) toxic bug report issues were linked with a PR that is lower than the percentage reported by Li \textit{et} al.~\cite{li2018issue} (40.4\%) and de Souza \textit{et} al.~\cite{de2024revealing} (35.7\%). 
This discrepancy suggests that toxic bug threads are less likely to result in actionable fixes through PRs, indicating that they may be deprioritized or avoided altogether by the maintainers. Consequently, toxic threads are less likely to be addressed and more likely to remain unresolved, perpetuating frustration and dissatisfaction among users. For example, in one of the threads that was not linked to PRs, a collaborator provided a workaround solution, and it was promptly rejected by a user - \textcolor{red}{``No. It's not. It's sloppiness and incompetence. And it's even less acceptable when it's actively used as an excuse not to revert a broken commit. [...]''}~\cite{containers-podman-issues-22190}.

\smallskip
\noindent
\textbf{VII. Subsequent Toxic Comments}: In 24/81 (29.63\%) cases, we observed subsequent toxic comments following the initial instance of toxicity. Only 6/24 cases, the actual problem faced in the issue, was resolved. This phenomenon was particularly evident in discussions where the initial comments contained insults or entitlement. Such interactions often created a cascading effect, escalating the negativity and derailing the conversation. For example, as illustrated in Figure~\ref{fig:motivating-example}, a dismissive or aggressive remark led to further toxic exchanges, making it difficult to refocus on the original issue.

\smallskip
\noindent
\textbf{VIII. Code of Conduct Enforcement}: In 3 cases, the project maintainers enforced the Code of Conduct after toxicity occurred.



\vspace{-4pt}
\section{Implications and Future Directions}
\label{sec:impl}



We have several insights and recommendations based on our study.

\vspace{2pt}
\noindent
\textbf{Transparent and Automated Bug Severity and Priority Management System}: 
A robust automated system for managing bug severity and priority should ensure transparency and inclusivity by providing clear triage messages and consistent, visible labels such as ``high-priority," ``severity-high," or ``needs clarification." Automated explanations for deprioritized or closed bugs, coupled with features like user voting on severity and real-time updates, can align prioritization decisions with community needs while mitigating frustration.
Recent research underscores the value of integrating social features with technical analysis to enhance bug-related processes~\cite{huang2024bug, imran2024shedding, han2024characterizing}. A comprehensive system would leverage machine learning to predict severity, utilize sentiment analysis to assess urgency, and establish feedback loops for maintainers to explain decisions. Visual dashboards tracking bug resolution progress further promote transparency. These measures would foster trust, improve collaboration, and create a constructive environment for bug reporting while reducing the potential for toxicity.

\vspace{3pt}
\noindent
\textbf{Code of Conduct (CoC) for External Users: } An outsider from a project tends to use for toxicity than a member of the project. For that reason, project maintainers should apply the CoC when creating the issues for external participants of the project.

\vspace{3pt}
\noindent
\textbf{Existing Tool Improvement and New Tool Development}: While leveraging state-of-the-art technologies to identify toxic GitHub threads, we observed the presence of false positives, often caused by unique characteristics of bug reports, such as stack traces. Stack traces, crucial for diagnosing root causes, frequently include software engineering terminology that automated tools misinterpret as toxic.
For instance, terms like `.ass' or `nerd-font' in Stack Trace triggered false toxicity detections. Future tools should integrate domain-specific knowledge and context awareness to differentiate technical jargon from toxic language, ensuring more accurate detection without hindering productive discussions~\cite{imran2024shedding}.

\vspace{1pt}
\noindent
\textbf{A large-scale empirical study}: A potential research direction would be working on a large-scale study, which is necessary to provide more concrete and actionable recommendations for mitigating toxicity in bug report discussions. Such a study would enable the software engineering community to better understand the impact and associations of toxicity during the bug resolution process. 

\vspace{3pt}
\noindent
\textbf{Limitation:}
This study is limited to the analysis of 100 GitHub projects. Using the repository level and thread level criteria, we mitigate the external threat of selecting these repositories. We used ToxiCR~\cite{sarker2023automated} and LLaMA (\textit{LLaMA-3.1:70B})~\cite{touvron2023llama} for the selection of the toxic comments, which leads to construct validity. To mitigate this, we manually analyze whether the comments are toxic or not. However, we may miss some toxic comments that are not detected by each of the tools. Since we chose GitHub-based projects for analyzing the bug report toxicity, we may miss other software developers' platform where toxicity manifest differently. However, GitHub is the most popular and diverse platform for OSS, and thus, GitHub-based projects capture a wide range of toxic behaviors and diverse community interactions.






\vspace{-8pt}
\section{Conclusion and Future Work}
\label{sec:conclusion}

Toxicity in bug reports poses significant challenges to collaboration and negatively impacts bug resolution rates. Through a qualitative analysis of 81 bug threads, we observed that external members are more prone to toxic comments, often driven by frustration over perceived mismatches between bug severity and maintainers’ priorities. These insights emphasize the importance of addressing toxicity in project discussions to improve communication and efficiency. Our findings serve as a foundation for future work on developing automated systems to detect and manage toxicity in bug reports. Expanding this research through large-scale empirical studies and refining toxicity detection techniques remain essential goals for enhancing collaborative environments in software development.





\bibliographystyle{ACM-Reference-Format}  
\balance
\bibliography{references}

\end{sloppypar}
\end{document}